\documentclass[aps,pre,twocolumn,showpacs]{revtex4}
\usepackage{amssymb}
\usepackage{amsmath}
\usepackage{graphicx}

\begin{document}

\title[Collapse of dense PNIPAM brushes]{Density effects on collapse, compression and adhesion of thermoresponsive polymer brushes}

\author{Ibrahim B. Malham}
\author{Lionel Bureau}
\email{bureau@insp.jussieu.fr}

\affiliation{Institut des Nanosciences de Paris, UMR 7588 CNRS-Universit\'e Paris 6, 140 rue de Lourmel, 75015 Paris, France}


\begin{abstract}

We probe, using the Surface Forces Apparatus, the thermal response of poly(N-isopropylacrylamide) (PNIPAM) brushes of various grafting densities, grown from plasma-activated mica by means
of surface-initiated polymerization. We thus show that dense thermoresponsive brushes collapse gradually as temperature is increased, and that grafting density greatly affects their ability to swell: the swelling ratio 
of the brushes, which characterizes the thickness variation between the swollen and the collapsed state, is found to decrease from $\sim 7$ to $\sim 3$ as the number of grafted chains per unit area increases. 
Such a result, obtained
with an unprecedented resolution in grafting density, provides qualitative support to calculations by Mendez {\it et al.} [{\it Macromolecules} {\bf 2005} {\it 38}, 174]. We further show that, in contrast to swelling,
adhesion between two PNIPAM brushes appears to be rather insensitive to their molecular structure.

\end{abstract}

\maketitle



\section{Introduction}
\label{sec:intro}

Temperature sensitive polymers attract a growing interest for their use in designing `smart' surfaces or interfaces \cite{rev1,rev2,rev3}, which properties (wettability \cite{wet1}, adhesion \cite{adh1}, 
friction \cite{friction1,zauscher1},...) can be deeply altered by an external thermal 
stimulus. Such smart systems rely on the fact that thermosensitive polymers exhibit a lower critical solution temperature (LCST) when mixed with a solvent \cite{LCST1}. Below the LCST, polymer chains are 
well solvated and adopt a swollen coil conformation. As the temperature is increased across the LCST, the solvent quality goes from good to poor, phase separation occurs, and macromolecules collapse to display a dense
globular conformation. Such a behaviour is commonly encountered in liquid binary mixtures in which specific interactions ({\it e.g.} hydrogen bonding) exist between constituents \cite{hirsch,gold1}, as in aqueous 
solutions of 
poly(N-isopropylacrylamide) (PNIPAM), one of the most 
studied among thermo-sensitive systems \cite{pnipam1,pnipam2}. 

PNIPAM in water exhibits a LCST around 32$^{\circ}$C \cite{pnipam1,pnipam2}, above which polymer chains 
become hydrophobic. The coil-to-globule transition in bulk solutions occurs over a narrow range of temperature and
involves large variations in chain dimension\cite{pnipam2,pnipam3}. These properties, which have been recently exploited in microfluidics for the design of switches \cite{switch1} and valves \cite{valve1}
for flow control, have found their main applications in bio-engineering \cite{rev2,rev3,cellmicrofluid}. The use of PNIPAM coatings on solid surfaces is indeed
a widespread technique for cell adhesion control in tissue engineering \cite{okano1,okano2}, or for separation of biomolecules in temperature selective chromatography\cite{okano2}. 

Grafted PNIPAM brushes are viewed as promising surface modifiers in order to finely tune the interactions  between cells or proteins and the surfaces of biomedical devices\cite{okano2}.
However, empirical design of such thermoresponsive sustrates might reveal insufficient. This appears for instance from two recent studies on harvesting of cell cultures grown on thermoresponsive substrates,
that reached opposite conclusions about the influence of brush thickness on cell proliferation/detachment \cite{okano1,celladh}. 
This underlines
 that reliable control of interactions requires a 
deeper understanding 
of how the brush properties (surface energy, ability to swell/collapse, thermal response,...) depend on molecular parameters
such as grafting density or chain length \cite{rev3}.

The structure of PNIPAM brushes in water has been the subject of recent experimental investigations. Force sensing techniques (Atomic Force Microscopy and Surface Forces Apparatus) have been used to
measure (i) the range of repulsive forces resulting from brush compression \cite{biggs,afm1,zauscher2,leckband1,leckband2}, and (ii) adhesion forces between PNIPAM brushes and various countersurfaces 
\cite{afm1,adh1,zauscher2,leckband1,leckband2}. Most of these studies have been performed at only two temperatures, 
below and above the bulk LCST. On the other hand, neutron reflectivity \cite{lopez1}, quartz-cristal microbalance \cite{biggs,qcm2,qcm3} or 
surface-plasmon resonance \cite{lopez2} have been used to follow in detail the  evolution with temperature of the thickness of brushes of different densities and molecular weights. 
Bringing together the conclusions of these studies, two important results may be put forward:

(i) the collapse of densely grafted chains occurs over a broader temperature range \cite{lopez1,lopez2,qcm2,qcm3} than that of dilute chains in solution \cite{pnipam3} or brushes with low grafting 
density \cite{biggs},

(ii) the swelling ratio, {\it i.e.} the ratio of the swollen to the collapsed thickness, is affected by both chain length and grafting density \cite{lopez1,leckband1,leckband2}. 

The former result has been obtained in several independent studies, and is in qualitative agreement with numerical and theoretical predictions about the collapse of brushes in poor solvent \cite{binder,zhulina1,zhulina2}. However, open 
questions remain regarding point (ii), due to the limited number of studies which systematically investigated  the effect of molecular parameters on the magnitude of brush collapse \cite{lopez1,leckband1,leckband2}.
In particular, whether a high grafting density favors or prevents large swelling ratio is still to be clarified. Furthermore, the role of brush density on the magnitude of adhesion forces with PNIPAM surfaces has 
remained unexplored.

These points, which are crucial for the design of smart surfaces based on thermoresponsive thin films, have led us to investigate in further details the effect of grafting density on the properties of PNIPAM brushes. 

We have used the surface forces apparatus to study the thermal response of PNIPAM brushes grown on mica substrates by the grafting from method. 

We focus on brushes of given molecular weight 
($M_{w}\approx 475$ kg.mol$^{-1}$) exhibiting densities ranging from $2\times 10^{-4}$ to $4\times 10^{-3}$ chain.\AA$^{-2}$. We show that, in the range of temperatures explored (20--40$^{\circ}$C):

(i) for all grafting densities,  the brush thickness decreases slowly as the temperature is increased up to 30$^{\circ}$C, then exhibits a marker decrease between 30 and 35$^{\circ}$C, temperature above which the 
collapsed thickness is reached. This confirms the broadening of the collapse transition previously observed for dense brushes \cite{lopez1,lopez2,qcm2,qcm3}.

(ii) the swelling ratio decreases noticeably with increasing surface coverage. This trend, which we observe with an unprecedented resolution over a large range of grafting densities, is in good qualitative agreement with recent predictions 
of Mendez {\it et al.} \cite{mendezsim}. 

(iii) in contrast to the density dependence of collapse, adhesive forces between two identical brushes, which build up for temperature above $\sim 30^{\circ}$C, appear to be rather insensitive to grafting density.

\section{Materials and Methods}
\label{sec:exp}

\subsection{Materials}

Substrates were cleaved from muscovite mica plates purchased from JBG-Metafix (France). The monomer N-isopropylacrylamide 
(NIPAM 99\%, Acros Organics) was recrystallized three times in n-hexane and stored in a desiccator under 
vacuum until use. n-hexane, toluene and absolute ethanol (Normapur, VWR France) were used as received. 3-aminopropyl-triethoxysilane 
(APTES, $\geq$ 98\%) and propyl-trimethoxysilane (PTMS, 98\%) were obtained from Sigma-Aldrich. Triethylamine (TEA, 99\% pure), copper (I) bromide 
(CuBr 98\%, extra pure), copper (II) bromide (CuBr$_{2}$, 99\% extra pure), 1,1,7,7-Pentamethyldiethylenetriamine (PMDETA, 99\%) and 2-bromo-2-methylpropionyl bromide 
(98\%, pure) were purchased from Acros Organics. All organic solvents were filtered over 0.20 $\mu$m Teflon membranes prior to use. All aqueous solutions were 
prepared in deionized water (18 M$\Omega$) filtered over a 0.20 $\mu$m cellulose acetate membrane before use. 

\subsection{Surface Forces Apparatus}

\begin{figure}[htbp]
$$
\includegraphics[width=8cm]{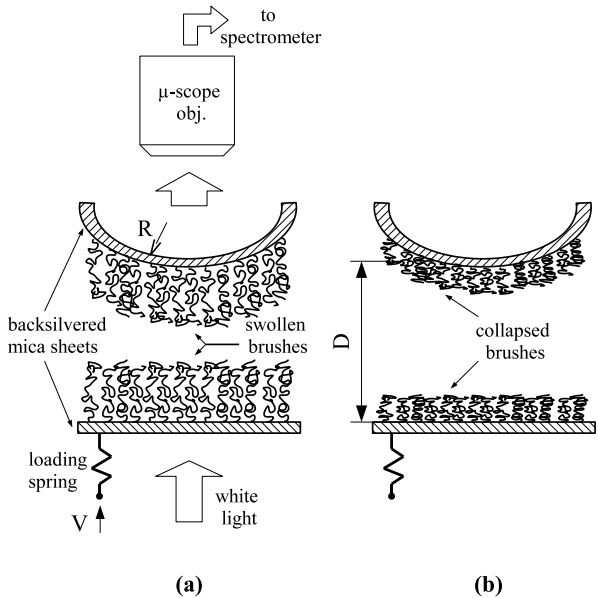}
$$
\caption{Schematic representation of the experimental configuration. PNIPAM brushes are grafted on curved mica sheets. Interbrush forces during compression/decompression are measured by means of a 
loading spring. The distance $D$ between the mica surfaces is measured by multiple-beam interferometry. The Fabry-Perot cavity formed by the backsilvered mica sheets is shone with white light. The transmitted 
light is collected by a microscope objective and sent to a spectrometer for spectral analysis of the fringes of equal chromatic order, from which $D$ is calculated. Experiments are performed over a temperature range of 
20--40$^{\circ}$C, which allows to probe thermal response of the brushes when going from the swollen (a) to the collapsed (b) state.}
\label{fig:fig1}
\end{figure}

Experiments reported in the present paper were performed using a home-built surface forces apparatus (SFA) \cite{rsi}. The apparatus has been modified to include a 
stainless steel cell of volume 1.5 mL that can be filled with a solvent into which the surfaces under study are fully immersed. We have used this instrument for time-resolved
measurements of force-thickness curves during compression/decompression of PNIPAM brushes in water, confined between two atomically smooth mica surfaces. Such measurements were performed at different
temperatures, below and above the bulk collapse temperature. The experimental configuration, involving two facing brushes grafted on curved mica sheets, is sketched on Fig. \ref{fig:fig1}.

Normal forces are measured, with a resolution of
$\sim10^{-6}$ N, by means of a flexure-hinge spring of
 stiffness 9500 N.m$^{-1}$ equiped with a capacitive displacement sensor. The distance between mica substrates is measured by white-light multiple-beam interferometry (MBI) \cite{sfa1,sfa2}: the fringes
 of equal chromatic order produced between the reflective backsides of the substrates are analyzed using the multilayer matrix method \cite{heuberger,rsi} in order to deduce the thickness and the effective refractive index 
 of the confined medium.
 
 Mica substrates were prepared according to the following protocol. A large mica sheet (10 cm $\times$ 10 cm) was cleaved down to a thickness of $\sim$10--30 $\mu$m. 
One side was thermally evaporated with a 40 nm-thick silver layer, in order to obtain a highly reflective surface for multiple beam interferometry.
Samples of $\sim 1$ cm$^{2}$ were cut from this sheet with surgical scissors, and glued, silver side down, onto glass cylindrical lenses (radius of curvature $R\simeq 1$ cm), using a UV-curing glue (NOA 81, Norland).
Two glued sheets were then re-cleaved using adhesive tape, down to a thickness of 1--4 $\mu$m. The cylindrical lenses were then mounted inside the SFA, with their axis crossed at right angle, and the exact thickness 
of each substrate was determined by MBI, following a procedure described in \cite{pre78}. Samples were then unmounted and subjected to the various surface treatments described below in order to obtain brushes of PNIPAM 
covalently grafted on the substrates.

After grafting on both mica sheets, samples were installed back in the SFA, brushes were brought into contact in the absence of solvent, and the total thickness, $h_{\text{tot}}$, of grafted PNIPAM was measured. In 
what follows, we report the dry thickness of the various brushes as being $h_{\text{dry}}=h_{\text{tot}}/2$, assuming that an identical amount of polymer is grafted on 
each surface. The refractive index for such dry layers was measured to be 1.47$\pm 0.02$, in good agreement with values reported in other study\cite{leckband1}.

The surfaces were then separated, and water injected into the liquid cell until full immersion of the brushes. The temperature inside the SFA was set, to within $\pm 0.01^{\circ}$C, to a chosen value in the range 20--40$^{\circ}$C, by means of a thermoregulation system described previously \cite{rsi}. The surfaces were approached by driving the remote point of the force-measuring spring at a 
prescribed velocity $V$, until a normal load of $\sim 10^{-2}$ N was reached, and then separated at the same driving velocity. All the results presented below have been obtained with 
$V\simeq 4$ nm.s$^{-1}$. We have checked that lowering the driving velocity by one order of magnitude did not affect the range of measurable repulsive forces by more than a few percents. This indicates
 that, with the chosen approach velocity, forces mostly arise from steric repulsion between brushes and not from hydrodynamic drag due to solvent flow.
Besides, we have found that the effective refraction index ($n_{\text{eff}}$) for the brush/water system increases from $\sim 1.33$ for large (several microns) intersurface separations, up to
 1.43$\pm 0.02$ at the onset of repulsion between brushes. Under stronger brush compressions, $n_{\text{eff}}$ stays roughly constant, within experimental resolution, at a value of $\simeq 1.45$. 
 Since we focus, in what follows, 
 on brush compression, we present results for which the brush thickness has been calculated using a constant $n_{\text{eff}}= 1.45$.

\subsection{Substrates preparation}

\begin{figure}[htbp]
$$
\includegraphics[width=8cm]{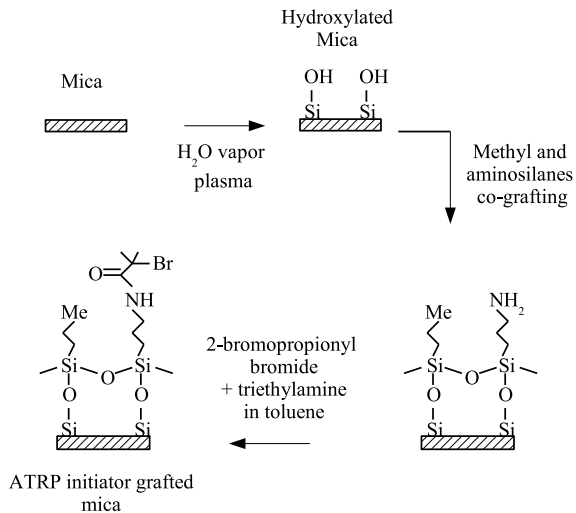}
$$
\caption{Grafting steps for immobilization of ATRP initiator on mica surfaces.}
\label{scheme1}
\end{figure}

Scheme \ref{scheme1} summarizes the different steps, following the thickness measurement of a pair of bare mica sheets, which lead to substrates grafted with a given surface density of initiator for Atom Transfer 
Radical Polymerization (ATRP).

(i) The mica samples of known thickness were transfered to a plasma reactor (model Femto from Diener Electronics Germany, operated at 80W), where they were exposed for 10 minutes to a RF-plasma generated in water vapor, at a pressure of 
0.4 mbar. Such a treatment, proposed initially by Parker {\it et al.} \cite{parker}, has been shown to produce silanol groups at the mica surface \cite{giasson1}, without any significant increase of  its roughness. 
Covalent attachment of silane based molecules on such plasma-activated surfaces is then possible \cite{parker,giasson1,giasson2,bureau1}.

(ii) Hydroxylated mica surfaces were then immediately immersed in a solution of propyltrimethoxysilane (PTMS) in toluene at 70$^{\circ}$C. The PTMS solution was prepared following the procedure described 
by Smith {\it et al.} \cite{smith}: 0.5 mL of PTMS were mixed with 25 mL of toluene, stirred at 70$^{\circ}$C for 90 minutes, and filtered over a 0.2 $\mu$m Teflon membrane before immersion of the substrates.
After PTMS grafting, mica surfaces were thouroughly rinsed with toluene, ethanol and water, and dried in an argon flow. This step results in a fraction of the silanol groups at the mica surfaces being replaced by methyl-terminated
molecules. This fraction was varied by adjusting the immersion time between 0 and 1h.

(iii) PTMS grafted samples were subsequently immersed, at room temperature and for a duration ranging from a few seconds to 5 minutes, in an aqueous solution of aminopropyltriethoxysilane (APTES). 
APTES (100 $\mu$L) was diluted in water (100 mL), stirred for 2h at ambient 
temperature in order to prehydrolyze the ethoxysilane groups, and filtered through a 0.2 $\mu$m membrane before use. APTES grafting was followed by copious rinsing with water and ethanol.
This yields mica substrates which exhibit a mixed NH$_{2}$/CH$_{3}$ functionality at the surface, the ratio of amino to methyl groups being varied by adjusting immersion times in step (ii)
and (iii).
 
(iv) A solution of triethylamine (1.22 mL) in toluene (20 mL) was then prepared and cooled at 0$^{\circ}$C. The silane grafted mica samples were immersed in this solution, to which 255 $\mu$L of 
2-bromo-2-methylpropionyl bromide were added dropwise. The samples were left in the bath for 2--3 minutes, after which they were rinsed in toluene, washed in ethanol and water and dried in an argon flow.
This step results in surface immobilization, on the amino-terminated sites, of the ATRP intiator from which PNIPAM brushes are grown \cite{lopez1,zhang}.

\subsection{Grafting of PNIPAM brushes}

Brushes  were grown by ATRP, following a protocole akin to that described in references \cite{lopez1,leckband1,zauscher2}.
A solution of NIPAM (3.00 g) dissolved in deionized 
water (15 mL) was deoxygenated by bubbling Argon at room temperature for 30 min. A solution of CuBr (0.044g) in water (3 mL) was prepared and stirred while 210 $\mu$L of PMDETA was added
 for copper complexation. This solution was then mixed to the NIPAM solution under argon bubbling. 

A pair of indentically modified mica samples
was then placed for 5 minutes in the aqueous solution of NIPAM and copper catalyst. 
Next, samples were immersed for 10 min in 10 mL of an aqueous solution of CuBr$_{2}$ (0.048g)/PMDETA (126 $\mu$L) 
to quench the polymerisation reaction. Finally, surfaces were rinsed repeatedly with water and dried carefully in an argon stream inside a laminar flow cabinet, where the were installed in the surface forces apparatus.

We have used the same polymerization time for each pair of mica substrates, which we expect to result in brushes formed of chains of comparable molecular weight, the surface density of which shall be controlled by the
initial ATRP initiator coverage. 

\subsection{Characterization of grafted layers}

PNIPAM brushes were first characterized by measuring their dry thickness, using the SFA. In what follows, we present results obtained on brushes of six different thicknesses, which are reported in Table \ref{table1}.
Since NIPAM polymerization occured only at the surface of the samples, and not in the bulk solution, we do not know {\it a priori} the molecular weight of the grafted chains. However, as presented in the Discussion
section below, a scaling analysis of the ratio between swollen and dry thicknesses, measured for each brush, allows us to estimate both the grafting densities and the chain length of the 
grafted layers. These values, reported in Table \ref{table1}, confirm that brushes of different thickness display different grafting densities of chains of roughly constant molecular weight.

\begin{table}
  \caption{Molecular parameters of PNIPAM brushes. $^{(a)}$estimated from the scaling analysis using Eq. 2. $^{(b)}$estimated from Eq. 1 after determination of the grafting density. $^{(c)}$density calculated 
  using Eq.1, assuming $N=4200$, {\it i.e.} the average value of chain lengths determined for the thicker brushes.}
  \label{table1}
  \begin{tabular}{lll}
    \hline
    dry & grafting density &  number of monomer\\
    thickness (nm) &  (chain.\AA$^{-2}$) &  units/chain \\
    \hline
    215 & 0.0042$^{a}$ & 4130$^{b}$ \\ 
    155 & 0.0033$^{a}$ & 3800$^{b}$ \\
    125 & 0.0024$^{a}$ & 4200$^{b}$ \\
    70 & 0.0014$^{a}$ & 4080$^{b}$ \\
    50 & 0.0009$^{a}$ & 4600$^{b}$ \\
    10 & 0.0002$^{c}$ & 4200$^{c}$ \\
    \hline
  \end{tabular}
\end{table}

Besides, we have performed wetting experiments in order to measure the static contact angle ($\theta_{s}$) of water as a function of temperature on the various brushes. This was done on flat mica substrates submitted, 
simultaneously 
with the SFA samples, to the sequence of chemical modifications described above. Experiments were performed using a custom-built instrument, described in a previous publication \cite{bureau1}, which 
we have adapted in order to 
regulate the temperature of the substrate and of the water droplets in the range 20--40$^{\circ}$C. 

\begin{figure}[htbp]
$$
\includegraphics[width=8cm]{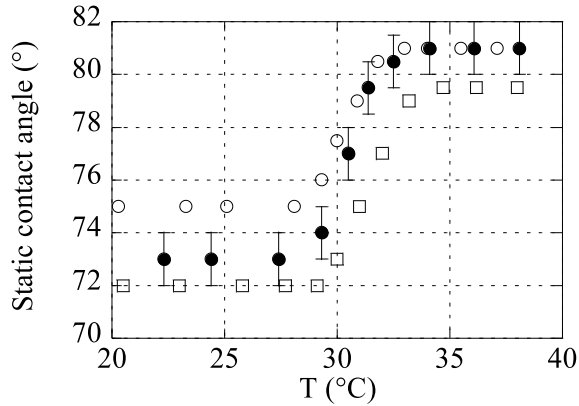}
$$
\caption{Static contact angle of water as a function of temperature, measured on PNIPAM brushes of thickness: ($\square$) 10 nm, ($\bullet$) 125 nm, ($\circ$) 255 nm. Error bars give the resolution on angle
determination, comparable to the dispersion of the results observed when measuring at various spots on a surface.}
\label{fig:fig2}
\end{figure}

Figure \ref{fig:fig2} shows the evolution of $\theta_{s}$ with temperature for three brush thicknesses. It can be seen that the contact angle stays constant for $T\lesssim 30^{\circ}$C, increases by approximately 
6--7$^{\circ}$ while $T$ increases from 30 to 34$^{\circ}$C, and exhibits a plateau at higher temperatures. Such a behavior is in qualitative agreement with previous measurements on similar surfaces 
\cite{lopez1,leckband1}. It shows that PNIPAM is indeed grafted on the mica substrates, the sharp increase of $\theta_{s}$ around 32$^{\circ}$C being the signature of the hydrophobic nature of PNIPAM above its 
LCST. As seen in Fig. \ref{fig:fig2}, we only observe a very weak influence of brush thickness on $\theta_{s}$. This  indicates that wetting is mainly sensitive to the chemical composition of the outermost region 
of the brush, and not to its detailed structure, 
as already suggested in previous works \cite{lopez1}.

\section{Results}
\label{sec:res}

\begin{figure}[htbp]
$$
\includegraphics[width=8cm]{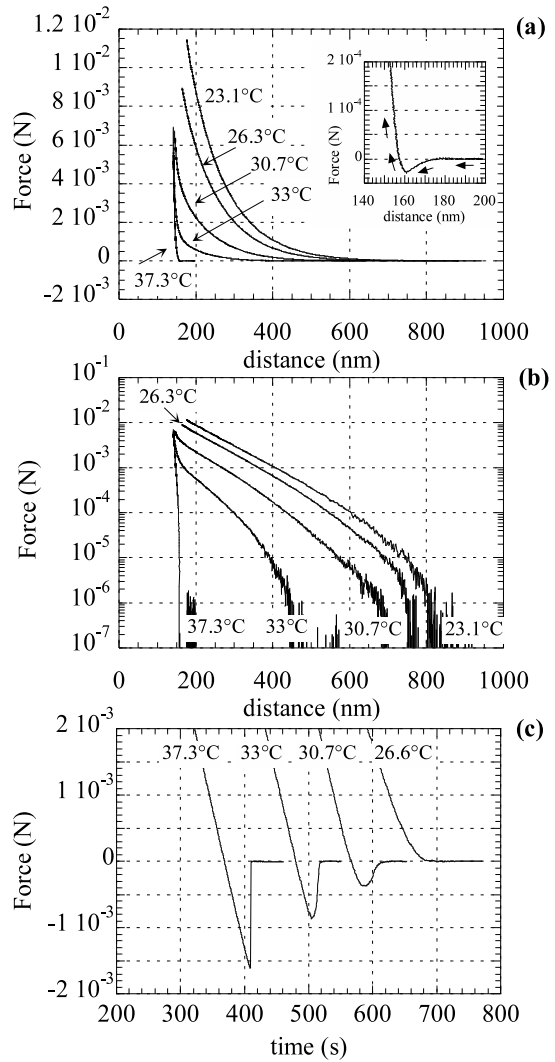}
$$
\caption{(a) Force (F)-distance (D) curves measured during approach of two brushes of dry thickness 70 nm, at the temperatures indicated on the figure. Inset: close-up of the $F(D)$ curve at 37.3$^{\circ}$C, showing
attractive interaction upon approach. (b) same data as in (a), with forces displayed on a logarithmic scale. (c) Force as a function of time, during separation of two brushes of dry thickness 70 nm, measured at the 
indicated temperatures. The curves have been horizontally shifted in order to the facilitate comparison.}
\label{fig:fig3}
\end{figure}

We first present the effect of temperature on the measured force-thickness curves during compression of two identical brushes. The results reported in Fig. \ref{fig:fig3} have obtained with 70nm-thick brushes.
The behaviors observed with thicker or thinner brushes displayed the same qualitative features. It can be seen  that, as the temperature is increased from 23 to 37$^{\circ}$C:

(i) the onset of repulsive forces occurs at smaller intersurface separations (Fig. \ref{fig:fig3}a and \ref{fig:fig3}b),

(ii) full compression of the PNIPAM 
layers (down to their dry thickness) is reached under lower normal forces (Fig. \ref{fig:fig3}a and \ref{fig:fig3}b),

(iii) upon surface separation, adhesive forces build up for $T\gtrsim 30^{\circ}$C, and are larger at higher temperatures (Fig. \ref{fig:fig3}c).

We also notice that at $T\gtrsim 35^{\circ}$C, in contrast to the monotonic repulsion measured at lower temperatures, attractive interactions between brushes are observed upon approach, before 
entering the `hard-wall' repulsion regime.
This is illustrated in the inset of  Fig. \ref{fig:fig3}a.

\begin{figure}[htbp]
$$
\includegraphics[width=8cm]{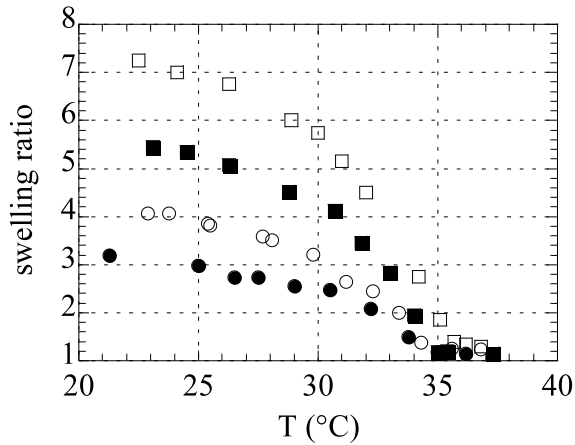}
$$
\caption{Swelling ratio $\alpha$ as a function of temperature, for brushes of thickness: ($\bullet$) 215 nm, ($\circ$) 125 nm, ($\blacksquare$) 70 nm, ($\square$) 10 nm.}
\label{fig:fig4}
\end{figure}

The effect of temperature on the range of repulsive forces is further characterized as follows. We measure, as a function of temperature, the thickness $h_{0}$ reached under a normal force of $10^{-5}$N, {\it i.e.}
under low compression. Note that the choice of such a criterion is arbitrary, it only ensures that the applied force is well above the experimental resolution, and $h_{0}$ is therefore lower than the unperturbed brush
thickness \cite{footnote1}. We define the ratio $\alpha=h_{0}/h_{\text{dry}}$, which we call the ``swelling ratio'' in the rest of the article. Fig. \ref{fig:fig4} presents the evolution of $\alpha$ with temperature, for 
different brush densities. We observe that:

(i) for all densities, $\alpha$ decreases gradually as $T$ increases, displays a stronger temperature sensitivity in the range 30--35$^{\circ}$C, and reaches, for $T\gtrsim 35^{\circ}$C, a constant value close to unity.,

(ii) the lower the brush density, the larger the value of $\alpha$ at low temepratures. 

\begin{figure}[htbp]
$$
\includegraphics[width=8cm]{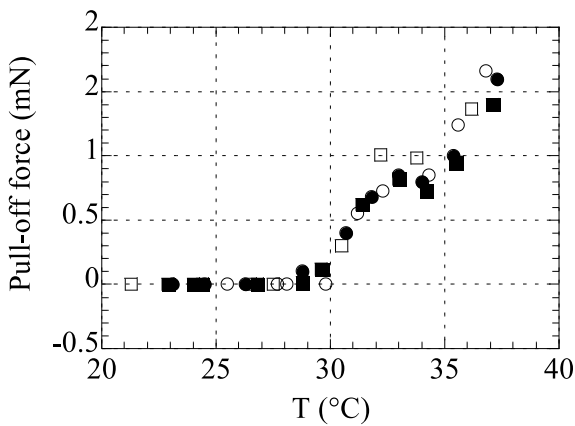}
$$
\caption{Pull-off force as a function of temperature, for brushes of thickness: ($\bullet$) 70 nm, ($\circ$) 125 nm, ($\blacksquare$) 155 nm, ($\square$) 215 nm.}
\label{fig:fig5}
\end{figure}

Finally, we plot on Fig. \ref{fig:fig5} the measured pull-off forces (the force needed to separate the brushes), as a function of temperature, for various brush densities.
It can be seen that adhesion forces between contacting brushes are independent of their grafting density, and increase with temperature above 30$^{\circ}$C.

\section{Discussion}
\label{sec:disc}

\subsection{Collapse}

The temperature dependence of the repulsive range reported in Fig \ref{fig:fig3}a and \ref{fig:fig3}b shows that the thickness of the brushes decreases as the temperature is increased: grafted chains evolve from
a low temperature swollen state (with $\alpha > 1$, see Fig. \ref{fig:fig4}) to a high temperature collapsed state ($\alpha\simeq 1$) where most of the solvent is expelled from the polymer layers.
This result is fully consistent with the thermal response of grafted PNIPAM previous observed in experiments using AFM or SFA \cite{afm1,biggs,zauscher2,leckband1}. 

Furthermore, the evolution of the swelling ratio with temperature (Fig. \ref{fig:fig4}) clearly shows that the collapse of end-tethered chains, in the range of grafting 
densities explored here,  occurs gradually rather than through a sharp coil-globule transition as observed for isolated chains in solution\cite{pnipam2,pnipam3}. Such a behavior, which is observed for the first time 
by means of force measurements, is in good agreement with results obtained from neutron reflectivity\cite{lopez2}, surface plasmon resonance\cite{lopez1} or quartz-crystal microbalance\cite{qcm2,qcm3}.
It is qualitatively consistent with theoretical predictions by Zhulina {\it et al.} \cite{zhulina1,zhulina2}, who have shown that, due to repulsive interactions between densely grafted stretched chains, 
the collapse of a dense planar brush occurs gradually as the solvent strength is lowered.

Besides, it can be seen on Fig. \ref{fig:fig4} that the grafting density affects both the magnitude of swelling and the  temperature sensitivity (the slope $\text{d}\alpha / \text{d} T$) of the brushes:
 as $T$ increases from 30 to 35$^{\circ}$C, $\alpha$ exhibits a 
sixfold decrease for the lowest density, whereas it drops by only a factor of 2.5 for the highest density. Such a result
is of importance for applications where large and sharp thickness changes are needed, {\it e.g. }  for efficient flow control or actuation in brush-grafted nano/microchannels \cite{switch1,cellmicrofluid}.

\subsection{Low temperature swelling}

\begin{figure}[htbp]
$$
\includegraphics[width=8cm]{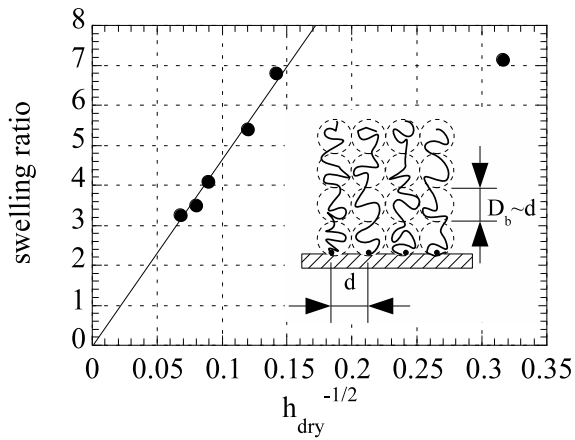}
$$
\caption{Ratio $\alpha$ as a function of $1/\sqrt{h_{\text{dry}}}$ (in nm$^{-1/2}$). The straight line is a guide for the eye. Inset: sketch of an Alexander-de Gennes brush made of blobs of size $D_{\text{b}}\sim d$.}
\label{fig:fig6}
\end{figure}

The low temperature values of the swelling ratio indicate that under good 
solvent conditions, less dense brushes swell more than denser ones (Fig. \ref{fig:fig4}). This result can be interpreted as follows, within the Alexander-de Gennes  framework for the description of brushes in good 
solvent \cite{alex,pgg}.\\
The dry thickness $h_{\text{dry}}$ of a brush is related to the grafting density by:

\begin{equation}
\label{eq1}
h_{\text{dry}}d^{2}=Na^{3}
\end{equation}
where $d$ is the average distance between tethering sites ($1/d^{2}$ is the grafting density), $N$ is the number of monomer units per chain, and $a$ is the monomer size. 
Following Alexander and de Gennes, we picture a swollen brush as formed of stretched chains subdivided into blobs of size $D_{\text{b}}\sim d$ (see inset of Fig. \ref{fig:fig6}). The blob size is given by the following
expression:
\begin{equation}
\label{eq2}
D_{\text{b}}=g^{\nu}a
\end{equation}
with $g$ the number of monomers in a blob. We discuss below the value of the exponent $\nu$.\\
The swollen brush thickness is:
\begin{equation}
\label{eq3}
h_{\text{swell}}=n_{\text{b}}D_{\text{b}}
\end{equation}
with $n_{\text{b}}=N/g$ the number of blobs per chain. Replacing $D_{\text{b}}$ by $d$ in Eq. \ref{eq2} and \ref{eq3}, one gets for the swollen thickness:
\begin{equation}
\label{eq4}
h_{\text{swell}}=Na\left(\frac{d}{a}\right)^{1-1/\nu}
\end{equation}
The ratio of the swollen to the dry thickness thus reads:
\begin{equation}
\label{eq5}
\alpha=\frac{h_{\text{swell}}}{h_{\text{dry}}}=\left(\frac{d}{a}\right)^{3-1/\nu}
\end{equation}
Such a scaling is expected to break down at low enough grafting densities, {\it i.e.} at large $d$, where the swelling ratio most likely saturates at a value comparable to that of an isolated chain.

Under the assumption of semi-dilute brush ({\it i.e.} for an average monomer volume fraction $\phi$ which stays $\ll 1$), the exponent $\nu$ shall 
be close to 3/5, the blob size being governed by excluded volume interactions \cite{pgg}. At large monomer concentrations, for very dense brushes, it has been shown that $\nu=3/5$ does not allow to
account for the measured swollen thickness\cite{pmmabrush}, and the evolution of $h_{\text{swell}}$ with grafting density observed in reference \cite{pmmabrush} suggests that
 high density brushes shall be better described with gaussian blobs, {\it i.e.} $\nu\simeq 1/2$.

On figure \ref{fig:fig6}, we plot $\alpha$, measured at 23$^{\circ}$C, as a function of $1/\sqrt{h_{\text{dry}}}$ (which is proportional to $d$). It can be seen that, at small $d$, the swelling ratio 
increases quasi-linearly with the distance between grafting sites, then levels off at large $d$ as expected. The linear increase of $\alpha$ with $d$ observed at small $d$ is in good agreement with 
expression \ref{eq5}, provided that $\nu\simeq1/2$. Such a value of $\nu$, in agreement with previous measurements on dense brushes\cite{pmmabrush}, is consistent 
with the fact that the average monomer concentration in the most dense brushes is large. Indeed, the volume fraction, which
 can be estimated as $\phi\simeq 1/\alpha$, is found to range from 1/7 to 1/3, values that may be too high for the assumption of semi-dilute brush to strictly hold.

In view of the agreement obtained between the measured and predicted evolution of $\alpha (d)$ at small $d$, we go one step further and use Eq. \ref{eq5}, assuming the monomer size to be 
$a\simeq 5$~\AA \cite{biggs,lopez2}, in order to evaluate the grafting densities of the brushes for which the linear scaling of $\alpha(d)$ holds. We thus obtain grafting densities, reported in table \ref{table1}, 
ranging from $9\times 10^{-4}$ to $4\times 10^{-3}$ chains/\AA$^{2}$. Next, we check the consistency of our assumption of constant $N$ by estimating the number of monomers per chain using Eq. \ref{eq1} and the 
calculated values of $d$. The polymerization indices are reported in table \ref{table1}, where it can be seen that the average value of $N$ is 4200, with variations of $\pm 10\%$ from brush to brush. This corresponds to
an average molecular weight of the grafted chains of 475 kg.mol$^{-1}$. Finally, we use the average value of $N$ to calculate, from Eq. \ref{eq1}, the grafting density of the brush for which the above scaling analysis
fails. 

\begin{figure}[htbp]
$$
\includegraphics[width=8cm]{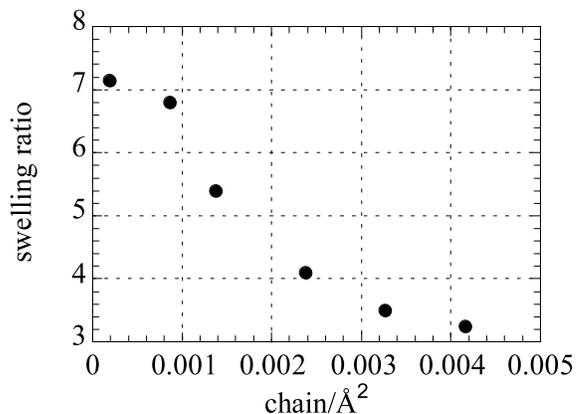}
$$
\caption{Swelling ratio $\alpha$ as a function of brush density.}
\label{fig:fig7}
\end{figure}

On Fig. \ref{fig:fig7}, we now plot the swelling ratios, measured at low temperature for the various brushes, as a function of the estimated grafting densities. Such a representation of the data allows for direct comparison 
of our results with those obtained in previous experimental and theoretical studies. We thus remark that, in a range of density similar to that covered in the work by Plunkett {\it et al.}\cite{leckband1}, we 
measure swelling ratios which are much larger than those reported in reference \cite{leckband1}, exhibiting moreover a density dependence of opposite sign with respect to that reported by Plunkett {\it et al.}. 
We have, at this
stage, no clear explanation for such a discrepancy \cite{footnote2}. However, we find that our data compares reasonably well with those of Yim {\it et al.}\cite{lopez2}, and confirm, on brushes made of longer chains, 
the general trend suggested by these authors. Furthermore, we find a striking qualitative similarity between the decrease of $\alpha$ with $1/d^{2}$  observed in Fig. \ref{fig:fig7}, and the 
prediction of Mendez {\it et al.}\cite{mendezsim} for dense brushes of long chains. 

\subsection{Compression of swollen brushes}
\label{subsec:force}

In an attempt to better identify the monomer volume fraction above which the assumption of semi-dilute brushes ceases to hold, we now discuss the validity of the Alexander-de Gennes framework for
 dense brushes under compression.

To this aim, we compare the low temperature compression curves, $F(D)/R$, with $R$ the radius of curvature of the surfaces \cite{footnote3,israelachvili}, with the following prediction for
the free energy per unit area of two compressed brushes \cite{alex,heuberger2,kleinbrush}:

\begin{equation}
\label{eq6}
\frac{F(D)}{R}=C\left[7\left(\frac{D}{2L}\right)^{-5/4}+5\left(\frac{D}{2L}\right)^{7/4}-12\right]
\end{equation}
with $L$ the unperturbed thickness of one brush, and the prefactor $C\propto k_{\text{B}}TL/d^{3}$, where $k_{\text{B}}$ is the Boltzmann constant and $d$ the distance between grafting sites.

\begin{figure}[htbp]
$$
\includegraphics[width=8cm]{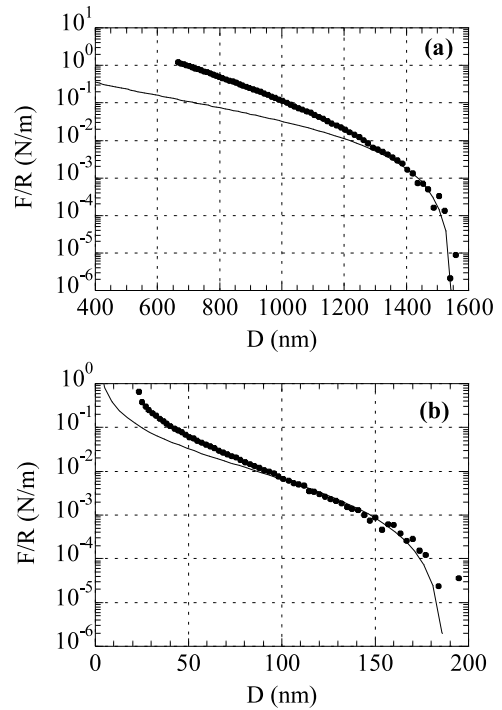}
$$
\caption{Normalized force $F/R$ as a function of brush thickness $D$ during compression of brushes of dry thickness: (a) 215 nm, (b) 10 nm. Symbols correspond to measurements, solid lines correspond to $F/R$
calculated with Eq. \ref{eq6}, using $2L=1550$ nm in (a) and $2L=190$ nm in (b).}
\label{fig:fig8}
\end{figure}

For the different brushes studied here, we have computed $F(D)/R$ using for $L$ the brush thickness measured at the onset of repulsion, and for $d$ the value estimated as described above. 
 Doing so, we find a partial agreement between predictions and measurements: the low compression end of the $F(D)/R$ profiles is reasonably well described by Eq. \ref{eq6}, but 
 the prediction  is found to systematically underestimate the repulsive forces under strong compression, as illustrated on Fig. \ref{fig:fig8}  \cite{footnote4}.
  Moreover, we observe that the higher the brush
 density, the narrower the range of agreement between experiments and theory: the calculated $F(D)/R$ fails to reproduce the 
 measurements for values of $D/2L$ lower than $\sim 0.85$ for the most dense brushes (Fig.\ref{fig:fig8}a), and for $D/2L\lesssim 0.6$ for the brushes of lowest density (Fig.\ref{fig:fig8}b).\\
 The existence of a compression  range over which agreement is fair may seem surprising, since we have shown in the previous section that $\alpha (d)$ was not well described under the assumption of a semi-dilute
 layer ($\nu\sim 1/2$ rather than 3/5). We interprete this apparent paradox as follows. Grafted brushes most likely exhibit, as already discussed in previous studies\cite{pgg,leckband1,heuberger2}, an outer region where the monomer 
 density is lower than that closer to
 the solid substrate ({\it i.e.} the monomer density profile is not step-like but rather decreases smoothly to zero). The above observations thus suggest that during compression of a dense brush, a first step consists in
 compressing this outer region of lower density, in which the semi-dilute assumption appears to hold, until reaching an average monomer density above which the brush is stiffer than predicted.

These results show that the Alexander-de Gennes model has only a narrow range of validity when dealing with dense brushes under compression. From the measurement of the thickness $D_\text{f}$ where predictions 
fail to account for the repulsive forces, one may estimate a monomer volume fraction $h_\text{dry}/D_{\text{f}}$, which we find in the range 0.2--0.3. We thus obtain, from force-distance measurements, an
estimate of the monomer density above which the assumption of semi-dilute brushes ceases to be valid.

\subsection{Wetting, attractive interactions and adhesion}

PNIPAM chains bear both hydrophobic (methyl -CH$_{3}$) and hydrophilic (amide -NH, and carbonyl -C=O) groups along their backbone.
As suggested by the temperature dependence of the water contact angle ($\theta_s$) shown on Fig. \ref{fig:fig2}, the outermost layer of a PNIPAM brush is expected to be the seat of local molecular rearrangements,
so that, at $T\geq 35^{\circ}$C, the hydrophobic groups are preferentially exposed at the interface with water. 
Besides, the range of temperatures over which $\theta_s$ increases roughly corresponds to the range where brush thicknesses exhibit a steeper decrease with $T$. 
This supports the idea that, as $T$ increases and chain dehydration and collapse proceed, a growing fraction of methyl groups is exposed at the surface, until a maximum surface density of CH$_3$ is reached 
above $35^{\circ}$C. 

Such a picture thus suggests that the attractive interaction between collapsed brushes observed at temperatures above $35^{\circ}$C (see Fig. \ref{fig:fig2}a and reference \cite{zauscher2}) 
most likely arises from the so-called hydrophobic forces that exist between methyl-terminated surfaces immersed in water \cite{christenson}. 

Now, the measurements of pull-off forces which do not depend on grafting density (Fig. \ref{fig:fig4}) indicate that adhesion between brushes in contact is, like wetting, mainly controlled by the nature of the 
chemical groups exposed at the interface. However, we note that while the water contact angle levels off above 35$^{\circ}$C, pull-off forces are found to increase steadily up to 38$^{\circ}$C. This suggests that,
once the contact is established between two collapsed PNIPAM layers, some mechanism is at play, which affects interbrush adhesion but not wetting. We believe that this mechanism corresponds to
the formation of interchain hydrogen bonds, between -NH and -C=O groups \cite{wu3} , thus contributing to increase the pull-off force. Such H-bonds are directional and require that the chemical groups 
involved are favorably oriented along the polymer chains. This should therefore result in a pull-off force that increases with the contact time, due to the dynamics of molecular rearrangements which
limits H-bonds formation. We are currently investigating this point, which constitutes a test of the above hypothesis of specific interactions at the interface between two pnipam brushes.

\section{Conclusions}

We have performed a systematic study of the effect of grafting density on the collapse, compression and adhesion of PNIPAM brushes. Our study, based on force/distance measurements in a Surface Forces Apparatus,
makes use of a plasma activation technique of mica surfaces that allows for covalent grafting of PNIPAM brushes by surface-initiated ATRP. We show that the thermal response of such brushes is strongly affected by the 
grafting density, in good agreement with recent theoretical predictions \cite{mendezsim}:  the swelling ratio is found to decrease noticeably 
at high densities, a result 
that may have direct implications in {\it e.g.} nano/microscale flow control using PNIPAM brushes \cite{switch1}. Furthermore, the analysis of force-compression curves  provides an insight 
regarding the range of validity of the classical Alexander-de Gennes framework when dealing with dense brushes.
Finally, we show that in contrast to the swelling behavior, adhesive forces between brushes do not seem to be affected by
the grafting density of the layers. Such an observation leaves completely open the question of how brush thickness may affect cell/substrate adhesive interactions in cell culture applications of PNIPAM \cite{okano1,celladh}.

\section{Aknowledgements}

We are indebted to Christiane Caroli for her contribution to data interpretation and her precious help with polymer brush theory.


\newpage









\begin{thebibliography}{achemso}
\bibitem{footnote3} Within the Derjaguin approximation, $F(D)/R$ is proportional to the interaction free energy per unit area \cite{israelachvili}.
\end{thebibliography}
\end{document}